\documentclass[preprint]{aastex}

\shortauthors{Gizis}
\shorttitle{TW Hya Brown Dwarfs}

\begin{document}

\title{Brown Dwarfs and the TW Hya Association}

\author{John E. Gizis\altaffilmark{1}}

\affil{Department of Physics and Astronomy, University of Delaware, 
Newark, DE 19716}

\altaffiltext{1}{Visiting Astronomer, Cerro Tololo Inter-American Observatory,
National Optical Astronomy Observatories, which is operated by
the Association of Universities for Research in Astronomy, Inc. 
(AURA) under cooperative agreement with the National Science Foundation.}

\begin{abstract}
I report the results of a survey for low-mass ($0.030 \gtrsim M \gtrsim 
0.013 M_\odot$)
brown dwarfs in the direction of the
TW Hya association using 2MASS. Two late-M dwarfs show signs of low surface
gravity and are strong candidates to be young, very-low-mass
($M \approx 0.025 M_\odot$) brown dwarfs related to the TW Hya association.
2MASSW J1207334-393254 is particularly notable for its strong 
H$\alpha$ emission.  The numbers of detected brown dwarfs is consistent with 
the substellar mass function in richer star formation environments.
Newly identified late-M and L dwarfs in the field
are also discussed.  Unusual objects include an L dwarf with strong H$\alpha$ 
emission, a possible wide M8/M9 triple system, and a possible L dwarf companion
to an LHS star.  
\end{abstract}


\section{Introduction}

The Two Micron All-Sky Survey (2MASS), the Deep Near-Infrared Survey (DENIS),
and the Sloan Digital Sky Survey (SDSS) have identified large numbers of 
cool dwarfs in the field.  This has led to the
development of the new L dwarf spectral class \citep{k99,m99},
corresponding to temperatures in the range $\sim 1400-2200$K.
The observed numbers of L dwarfs imply that  
isolated field brown dwarfs are common though even if the details of the 
substellar mass function remains uncertain \citep{reidmf}. 
Observations of rich star-forming regions have also identified numerous
young brown dwarf candidates (see the review of Basri 2000).   

Recent work has identified nearby ($d \lesssim 100$ pc) associations of 
very young ($<100$ Myr) stars \citep{youngstars}.   The most notable of
these new associations is the TW Hya association (TWA), 
a loose group of T Tauri stars with an age of $\sim 10$ Myr 
\citep{webb99,mf01}.  These nearby T Tauri and post-T Tauri stars 
allow unique studies of star formation processes due to their
proximity and brightness.  Furthermore, they offer insights into
star formation in a low-density enviroment that differs from 
the dominant star formation process 
observed in dense star regions like Orion. 

The identification of young brown dwarfs in TWA and similar associations 
promises to advance substellar astronomy 
--- confirming that isolated brown dwarfs form in a 
low-density star formation environment and providing bright, nearby 
prototypes for follow-up studies.  
Current optical and X-ray surveys were only sensitive
to ordinary stars ($M \gtrsim 0.2 M_\odot$).  
Nevertheless, one young TWA brown dwarf is
known:  TWA 5B is an M8 secondary identified by Hubble Space Telescope 
imaging about 100 A.U. from its primary star \citep{lowrance99}.     
The apparent magnitude, $H=12.1$, of 
TWA 5B is well above the 2MASS magnitude limits ($H \approx 15.1$ 
for signal-to-noise 10, Cutri et al. 2000);  
any similar, {\it isolated} TWA brown dwarfs
will therefore appear in the 2MASS survey.  Furthermore, TWA is in a relatively
clean field away from the Galactic plane and without background reddened
sources.  

I present the results of a 2MASS-based search for isolated, very-low-mass
brown dwarfs in the TW Hya association.  I present the search criteria
and follow-up spectroscopy in Section~\ref{data}.  The results 
are discussed in Section~\ref{discussion}.  

\section{Sample Selection and Spectroscopy\label{data}}

\subsection{Photometric Selection}

The 2MASS survey provides precise and accurate photometry
in the J, H, and K$_s$ bands.  \citet{k99} and \citet{gizis00} discuss
the selection of late-M and L dwarfs using the 2MASS surveys.
M dwarfs in the range M0-M7 have similar near-infrared
colors ($J-K_s \approx 0.9$) which makes it difficult to 
efficiently select M6 and M7 dwarfs.  In contrast, M8 
through late-L dwarfs are characterized by redder 2MASS colors 
($J-K_s>1.0$) which distinguish them from most background stars.
The addition of optical colors from the POSS and other 
optical surveys allow nearly all contaminating populations
(mostly extragalactic at high galactic latitude) to 
be eliminated.  I therefore chose to concentrate on these
M8 and cooler L dwarfs in this initial search for TWA candidates.
Because the hydrogen burning limit at this age will lie near spectral class M6
\citep{luhman}, this strategy allows only the cooler and hence
lower mass brown dwarfs to be detected, roughly corresponding
to masses $M<0.03 M_\odot$.  

\begin{figure}
\epsscale{0.7}
\plotone{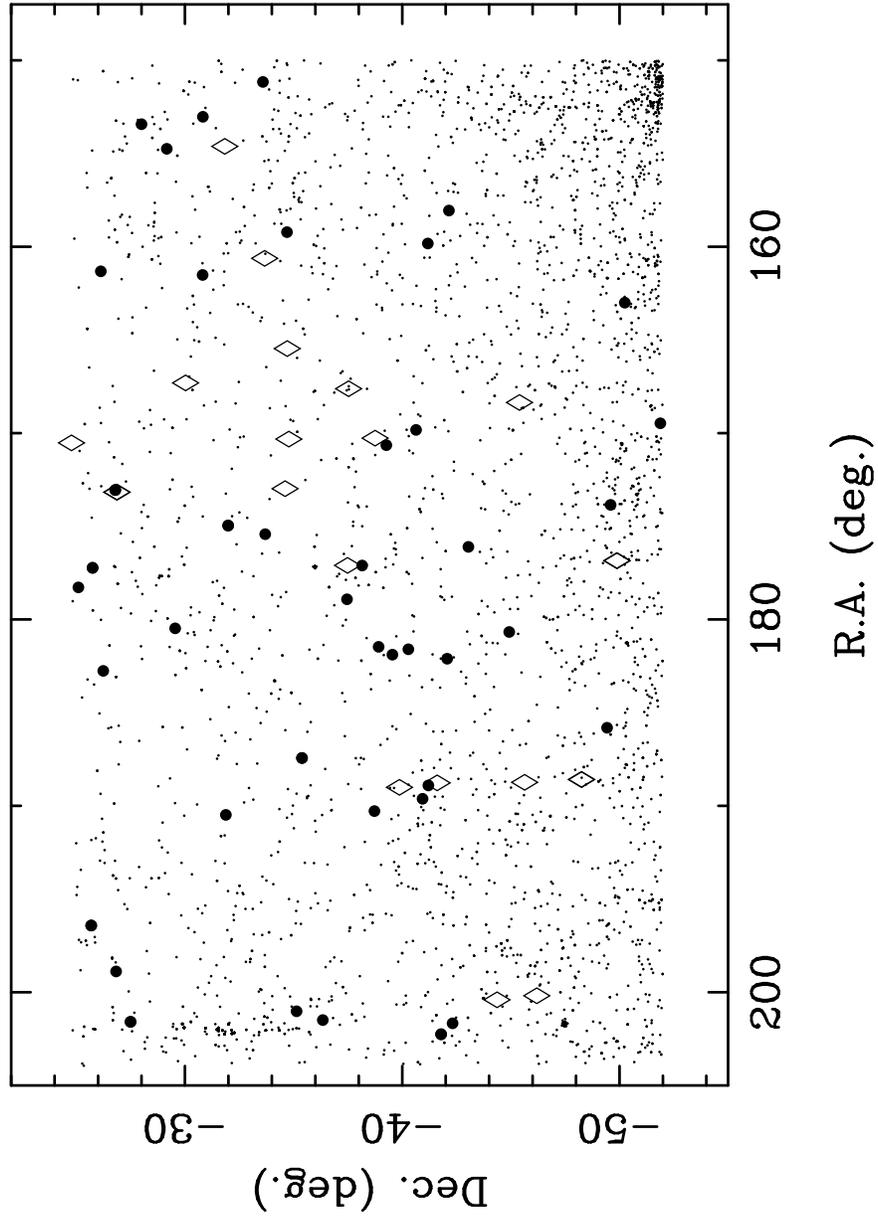}
\caption{Spatial distribution of known TWA members (open diamonds),
and candidate M/L dwarfs for this program (solid circles).
\label{fig-area}}
\end{figure}

\begin{figure}
\epsscale{0.7}
\plotone{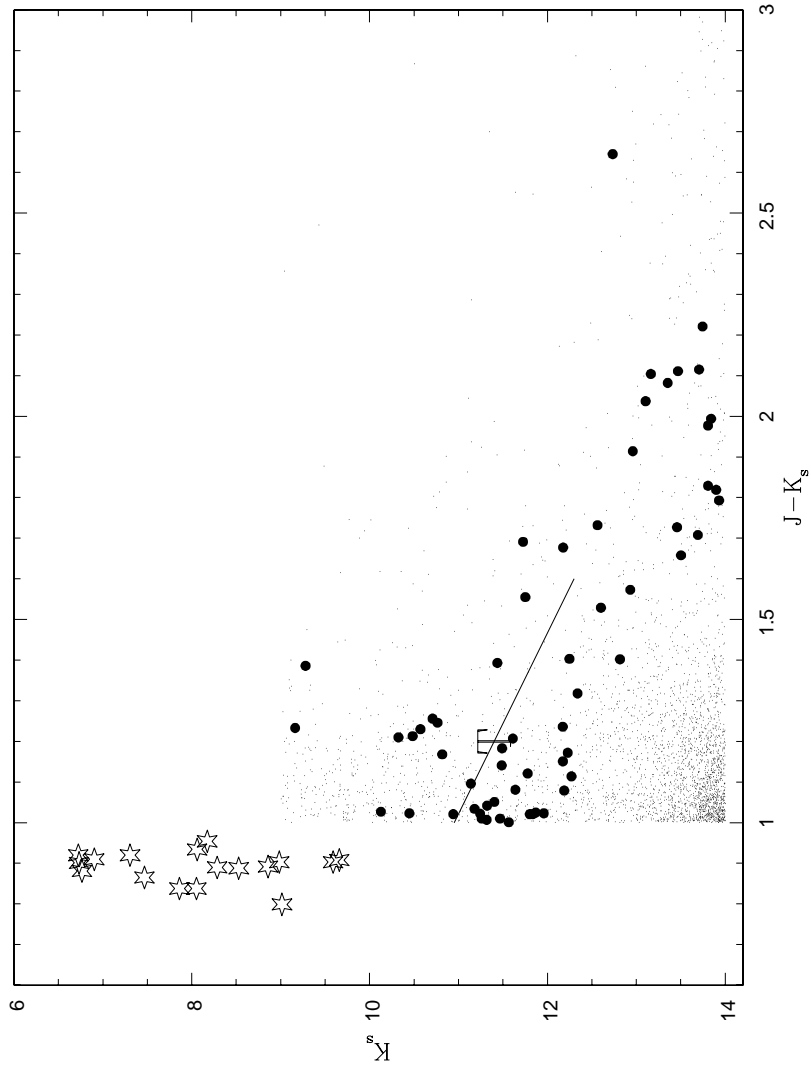}
\caption{The 2MASS color-magnitude diagram including 
known TWA members (open stars),
candidate M/L dwarfs for this program (solid circles), and 
other red background 2MASS sources (small points) with
optical counterparts.  The data for brown dwarf TWA 5B  
\citep{lowrance99} is marked as a large 'T.'  The solid line
plots the observed field M/L color-magnitude relation shifted
to agree to TWA 5B.  
\label{fig-colormag}}
\end{figure}

Figures~\ref{fig-area} and~\ref{fig-colormag} illustrate the
initial sample selection using the 2MASS Working Database.  
This database includes the entire sky but represents only the
preliminary processing of the 2MASS dataset.  
The area searched (Figure~\ref{fig-area}) 
was chosen to encompass the TWA members listed by 
\citet{webb99}, \citet{sterzik99} and \citet{zuck01}.  
The photometric selection (Figure~\ref{fig-colormag}) 
was designed
to include potential late-M and L dwarfs members on the
basis of their $J-K_s$ colors.  The position of TWA 5B
is marked in Figure~\ref{fig-colormag} as a T.  \citet{luhman} has found 
that the young brown dwarfs in IC 348 (age $\lesssim 10$ Myr) 
have J-H,H-K colors similar to the
field population.  
The limiting magnitude for targets was a function of J-K$_s$:
\begin{equation}
K_s < 2.5 \times (J-K_s) + 9.5
\end{equation}
As shown in Figure~\ref{fig-colormag}, this selection would
include any other M/L dwarfs that follow the 
observed field M/L dwarf sequence slope found
by \citet{gizis00} and extends 1.1 magnitudes fainter
at $K_s$ than TWA 5B.   This allows for the inclusion of 
brown dwarfs further away than TWA 5B, or the possibility
that TWA 5B is overluminous due to being an unresolved double.  
The limiting magnitude for the survey corresponds to 
$\sim 0.013 M_\odot$ assuming bolometric corrections similar
to field dwarfs \citep{dusty00} --- this limiting mass is of course
uncertain due to both theoretical and observational limitations.  
Each 2MASS source was examined on the DSS sky survey scans
at the CADC web site.
Those with bright optical counterparts ($B \lesssim 19$) were
excluded since M8 and later dwarfs will be red.   
The targets are listed in Table~\ref{table-twhya}.   The J2000
positions encoded in the official 2MASS names are accurate to an
arcsecond.  

\subsection{Other Field Dwarfs}

Late in the night, the TWA region was not accessible 
during the CTIO observing run.  I therefore selected 
candidate nearby field M and L dwarfs using the 2MASS Working
database.  The targets observed are listed in Table~\ref{table-field}.  
All targets are required to have $J-K_s > 1.0$ and no bright optical
counterpart.  
A number of other known field M and L dwarfs were observed, most
notably the recent discovery  DENIS-P J104814.7-395606.1 
\citep{denis}.  I find a spectral type of M8 for this star, slightly
earlier than the M9 previously estimated in the discovery paper 
from high resolution observations.
TVLM 868-53850, an M dwarf discovered by \citet{tinney}, is a 
near-infrared standard star \citep{persson,sergei} that has
not been previously observed spectroscopically.  It is an 
M5.0 with strong H$\alpha$ emission (7.0\AA).  A careful examination
of the 2MASS calibration observations of this star might reveal if 
the high chromospheric activity leads to any near-infrared variability.
The standards VB8,  LHS 2397a, LHS 2065, and 
2MASSW J1507476-162738 
as well as some earlier M dwarfs were 
also observed.  No flares were noted in these dwarfs. 
 
\subsection{Spectroscopy}

The targets were observed using the R-C spectrograph on the 
CTIO 4-meter Blanco telescope on UT Dates 30 April to 3 May 2001.  
Despite the faintness of the targets, even the L dwarfs
were visible in the acquisition camera.  The wavelength coverage was 
6430 to 9700 \AA~ with a FWHM resolution of $\sim 2.8$\AA.  Flux standards
from \citet{hamuy} were observed.  The spectral extraction and calibration
was made using standard IRAF routines.  

\begin{figure}
\epsscale{0.9}
\plotone{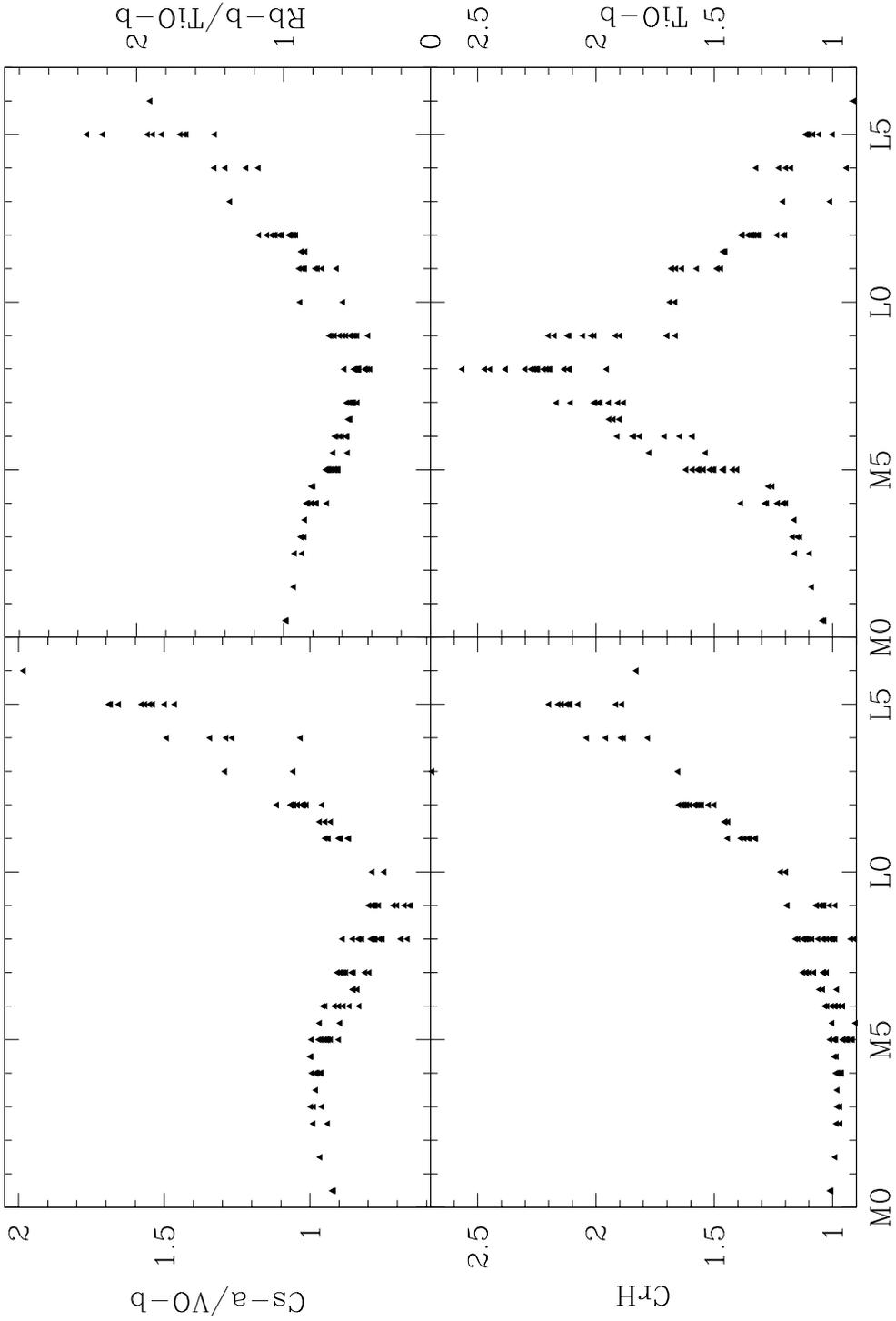}
\caption{Spectral indices plotted as a function of spectral type.  Each measurement of a star is plotted.
\label{fig-index}}
\end{figure}

To aid in spectral classification, spectral indices were measured.
These included TiO5 \citep{rhg95,g97}, PC3 \citep{m99}, 
CrH, TiO-b, Cs-a, VO-a and VO-b \citep{k99}.  
Spectral types were adopted based on
these indices and visual inspection of the spectra.  All spectral types
are on the \citet{khm91} and \citet{k99} M and L dwarf scale.  The equivalent
width of H$\alpha$ was measured when emission was present. 
The adopted spectral types and equivalent widths are included 
in Tables~\ref{table-twhya} and~\ref{table-field}.  The measured
indices are given in Table~\ref{table-indices}, where multiple measurements
of the same object have been averaged.  In 
Figure~\ref{fig-index},
some of the measured indices are plotted as a function of adopted
spectral type.  
The signal-to-noise and resolution of the spectra are not adequate to 
search for lithium.

\section{Discussion\label{discussion}}

\subsection{Two TW Hya Association Late-M Brown Dwarfs? \label{latem}}

\begin{figure}
\epsscale{0.9}
\plotone{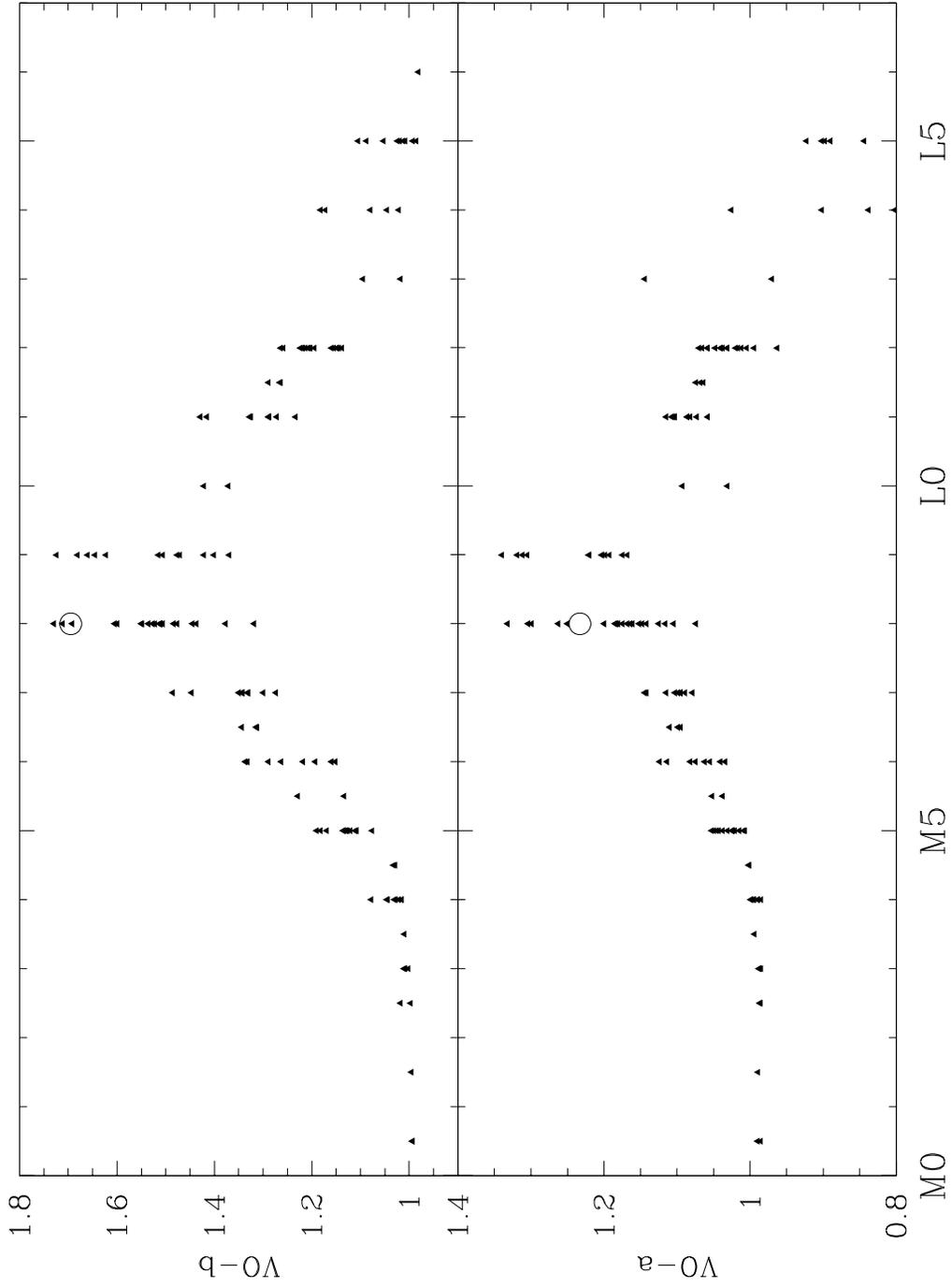}
\caption{Measured VO-a and VO-b indices  as a function of spectral
type.  \citet{luhman}'s brown dwarf is shown as an open circle.  Two 
dwarfs show evidence of strong VO and hence low surface gravity, both in
the TW Hya region.  Each measurement of a star is plotted.
\label{fig-sg}}
\end{figure}

Young brown dwarfs, particularly TWA brown dwarfs,
will have low surface gravity.  
\citet{llr97} and \citet{mrzo96} discuss the characteristics of
young late-M brown dwarfs.  They are characterized by stronger VO and 
weaker hydrides and Na lines than ordinary field M dwarfs.  
In Figure~\ref{fig-sg},
the surface gravity sensitive indices are plotted against 
spectral type.  
Only two late-M objects show both strong VO and 
weak Na and CaH.  This can be seen in Figure~\ref{figure-comp}, where the
specrtra of these two objects are compared to the ordinary field M8 dwarf
LHS 2397a and the \citet{llrl97} $\rho$ Oph brown dwarf.  
A few objects show an anomalous VO index but their other indices and the
overall appearance of their spectra do not support low surface gravity;
I attribute these anomalies to noise and do not consider them further.
The two candidate TW Hya brown dwarfs are discussed 
below.

{\bf 2MASSW J1207334-393254}:  This M8 dwarf shows signatures of
low surface gravity, which indicates that it is a young ($\lesssim 
100$ Myr) brown dwarf.  
Confirmation of TW Hya membership requires space motion information.
I have examined the United States Naval Observatory 
Image and Catalogue Archive for sky survey images of
2MASSW J1207334-393254.  
The oldest image, taken in 1964, has the poorest detection.  
The images show motion of 
100 mas~yr$^{-1}$ west and 30 mas~yr$^{-1}$ south relative to other
stars in the field; this motion is consistent with TWA membership
(Mamajek, private comm.).  
On the basis of this proper motion and the spectroscopy, 
2MASSW J1207334-393254 therefore may be regarded as a probable member
of the TW Hya association, although CCD astrometry is needed to
strengthen this conclusion. 

In any case, the proper motion is
considerably smaller than most 2MASS late-M dwarfs \citep{gizis00},
supporting a young age.

\begin{figure}
\epsscale{0.9}
\plotone{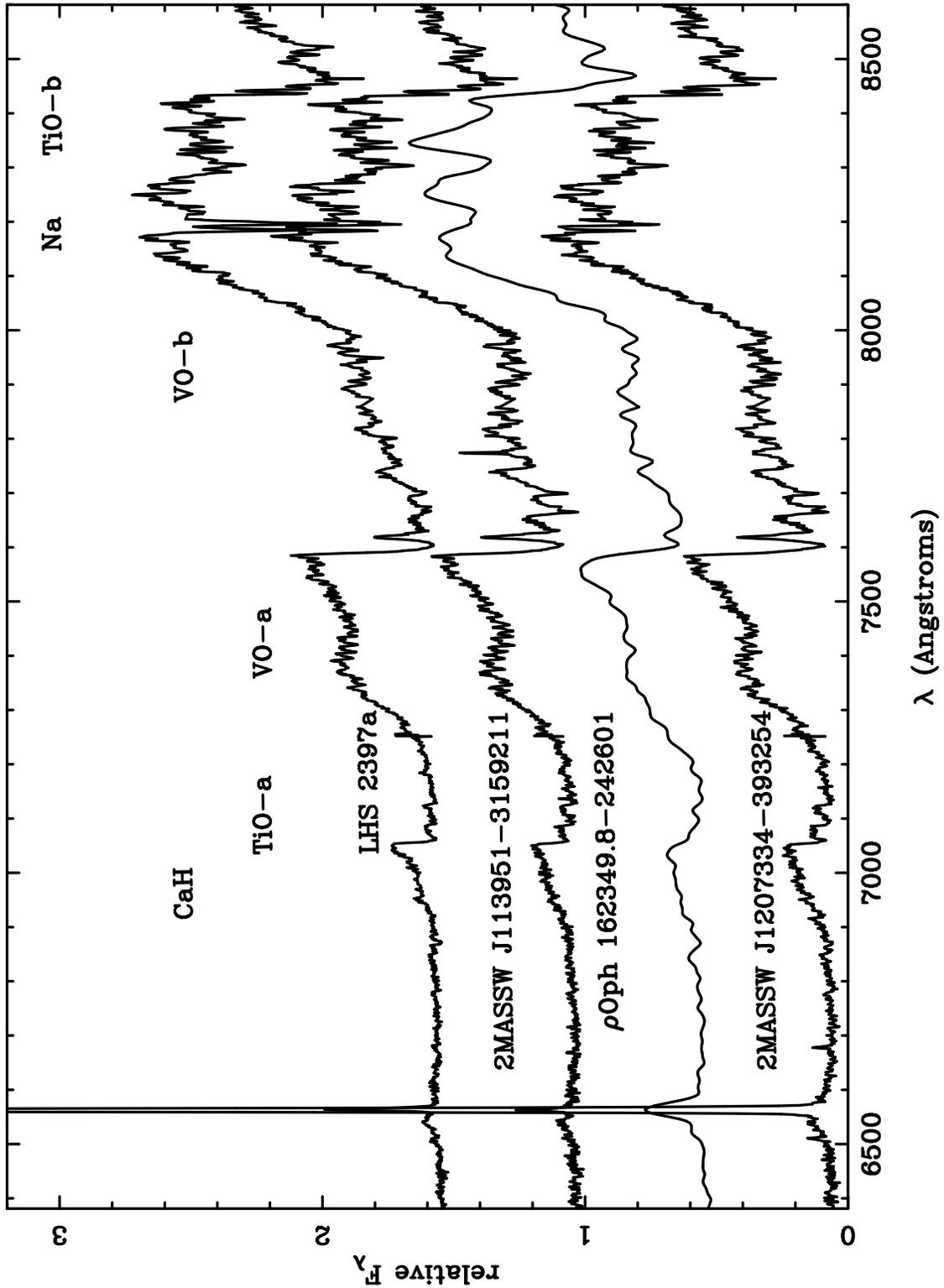}
\caption{\citet{luhman}'s $\rho$ Oph brown dwarf and the young brown dwarf
candidates  
2MASSW J1207334-393254 and 2MASSW J1139511-315921.  The field
M8 dwarf LHS 2397a is also shown.  Note the weak CaH, weak Na, and 
strong VO compared to the field dwarf.
\label{fig-comp}}
\end{figure}

2MASSW J1207334-393254's most notable characteristic, however, is its
extremely strong H$\alpha$ emission (Figure~\ref{fig-comp}). 
The equivalent width of H$\alpha$
emission in two consecutive 900 second exposures was $~300$\AA.
Helium emission
at 6678\AA~ is also detected with an equivalent width of 3\AA.
A powerful flare such as the one observed by \citet{liebertflare} could
account for the H$\alpha$ emission, but would have weakened by the
second exposure and produced additional atomic emission lines. 
Since these characteristics are not observed, it 
is unlikely that this emission is caused by a flare.  
  
Strong but persistent H$\alpha$ emission is not seen in field 2MASS late-M 
dwarfs \citep{k99,gizis00}, but is known in three late-M systems.
Two of these systems, the M8 brown dwarf $\rho$ Oph 162349.8-242601
\citep{llr97} and the M7 brown dwarf CFHT-BD-Tau J043947.3+260139 
\citep{mtaurus}, are young brown dwarfs whose emission is attributed
to an accretion disk.  
If due to the same cause, then 2MASSW J1207334-393254's
emission supports a young age and probable TW Hya membership.  
Figure~\ref{fig-comp}

Strong H$\alpha$ emission is also known in 
another, still mysterious, M9 dwarf.  
The dwarf PC~0025+0447 was discovered by \citet{pc0025discover} 
due to its H$\alpha$ emission ($EW \approx 300$\AA) which has persisted
for a decade.
2MASSW J120733-393254's H$\alpha$
and HeI emission are very similar to \citet{mbzo99}'s spectra
of PC~0025+0447 in its relatively inactive, unveiled state.
\citet{mbzo99} argue that PC0025+0447 is a young brown dwarf
with an age between 10 and 100 Myr and that the emission is due to 
chromospheric/coronal activity.  Such a scenario might also apply for
2MASSW J1207334-393254  and would be consistent with TWA membership.

{\bf 2MASSW J1139511-315921}:  
A second M8 dwarf, 2MASSW J1139511-315921, is also 
a low surface gravity brown dwarf and therefore a candidate
TWA member.  The $H\alpha$ emission is weaker, with 
$EW = 9.7$\AA, but He 6678\AA~ emission is present with  $EW \approx 1.9$\AA.
Weak emission is typical of young brown dwarfs.
Examination of the USNO scans reveals
a proper motion consistent with $\sim 110$ mas~yr$^{-1}$ to the south and
$\sim 110$ mas~yr$^{-1}$ to the west.  This motion is inconsistent with the
TWA and other young clusters, except perhaps the Pleiades (Mamajek, private
comm.).  Nevertheless, given the weak 
detections and resulting astrometric uncertainty 2MASSW J1139511-315921 may
still be a TWA member.  Comparison to the \citet{dusty00} models 
suggests that both 
2MASSW J1207334-393254 and 
2MASSW J1139511-315921 have masses $\sim 0.025 M_\odot$ if members;
\citet{lowrance99} finds the same result for the similar TWA 5B.  

\subsection{L Dwarf Candidates \label{twhya_ldwarf}}

Seven L dwarfs were detected in the TWA search.  
If any of these L dwarfs were TWA members, they would
have very low mass and low surface gravity.  
\citet{m99} have compared the L2 dwarf 
G196-3B, believed to have an age of $\sim 100$ Myr, 
with older field L dwarfs of similar spectral type.  It shows weaker
TiO, CrH, FeH, Na, K, Cs and Rb.  It is  
likely that $\sim 10$ Myr old early-L dwarfs will show 
similar, but greater, anomolies.  No later very low-surface gravity L dwarfs
are known, but it seems likely they would have also have significant 
differences with respect to ordinary field L dwarfs.
Since none of the observed
L dwarfs show compelling evidence for low surface gravity, I conclude they
are not TWA members, but three 
deserve special discussion.  Measurement of the proper motions of 
the L dwarfs would be relatively easy over the next few years and
would indicate if any are actually TWA members.  

\begin{figure}
\epsscale{0.9}
\plotone{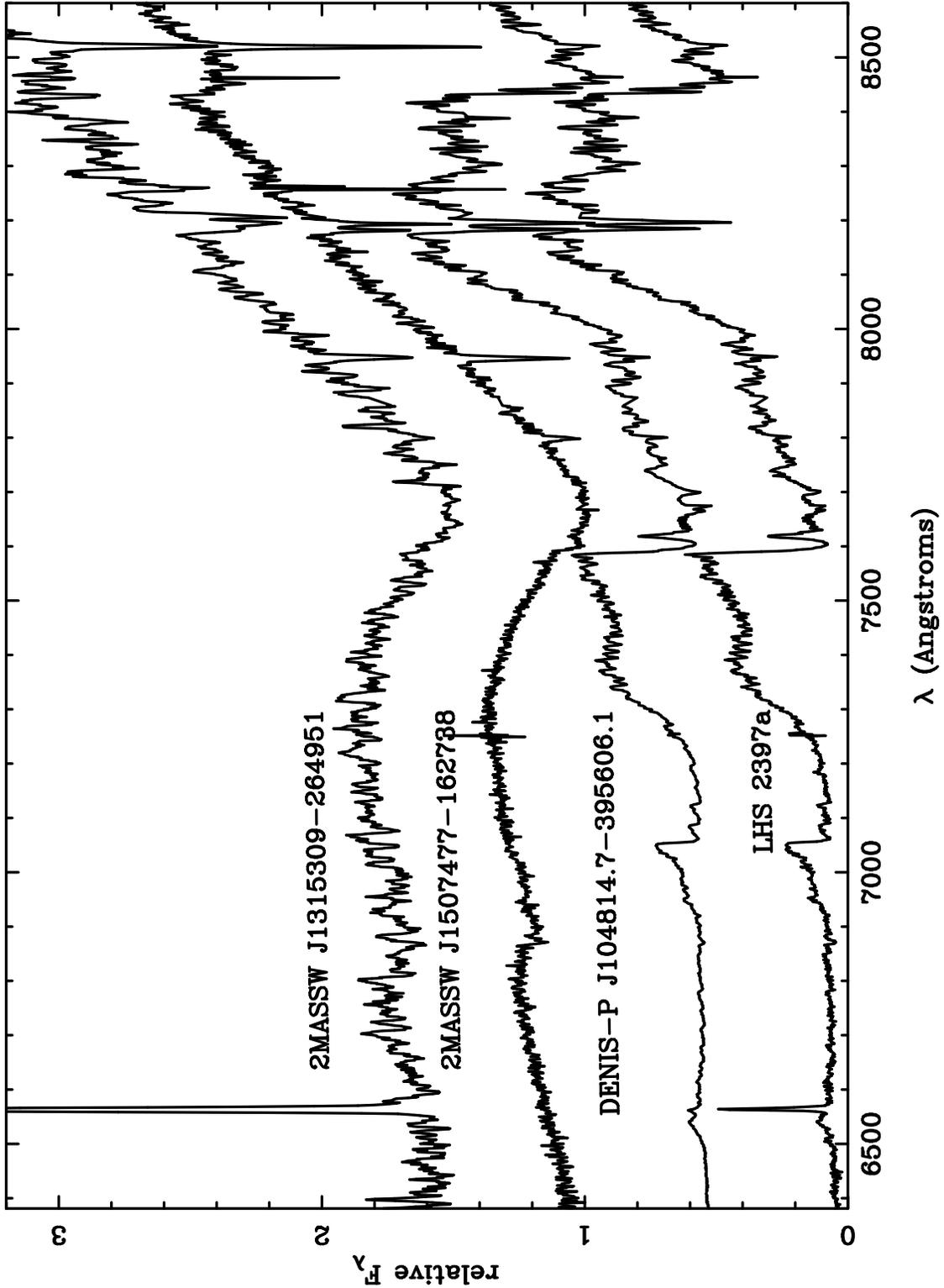}
\caption{The active L dwarf 2MASSI J1315309-264951 (boxcar smoothed) 
compared to the
field L5 dwarf 2MASSW J2MASSW J1507477-162738 and the
nearby M dwarf DENIS-P J104814.7-395606.1 compared 
to the M8 dwarf LHS 2397a.  
\label{fig-other}}
\end{figure}

{\bf 2MASSI J1315309-264951}:  
The L5 dwarf 2MASSI 1315309-264951 is an intriguing
object because it shows strong H$\alpha$ emission
(Figure~\ref{fig-other}).  Because of the
very low continuum levels, the equivalent width of $\sim 100$\AA~
is highly uncertain.  The relative energy in H$\alpha$ is 
$\log (H\alpha/L_{bol}) = -3.9$.  This L dwarf is $\sim 100$ times
more active than the {\it upper limits} for late-L dwarfs in 
\citet{gizis00}.  
It was observed with six consecutive
600 second exposures on 
01 May and two consecutive 900 second exposures
on 02 May.  The emission is constant over those time periods and 
it is therefore unlikely that the emission is due to a flare.
The J-K$_s$ color of 1.7 is not unusual 
for the spectral type of L5.  

2MASSI J1315309-264951 was independently discovered by \citet{hall01}, who
discribes two spectra.  \citet{hall01} found strong emission on 30 March 2001 with EW$\approx 100$\AA, but much weaker emission (EW $\approx 25$\AA) in a
lower-quality spectrum in August 2001.\footnote{Note that \citet{hall01} assigns a spectral type of L3, but the spectra 
reported are nearly identical to the L5 2MASSW J1507476-162738.}  
\citet{hall01} gives a discussion of the possible analogs and 
explanations for the emission.  Hall 
favors the explanation that the March emission is due to a flare.  Given the match of the emission in the May spectra, this now appears
to be an unlikely possibility.  Confirmation of the low activity state seen
in August is desirable.
Hall discusses the possibility that 2MASSI J1315309-264951 may 
be an analog to the Te dwarf 2MASSW 1237392+652615 
\citep{tedwarf}.    
Since, like the Te dwarf,  2MASSI J1315309-264951  
does not show signs of low surface gravity,   
it is likely a hotter analog of the still not-yet-
understood Te dwarf rather than a very young (TW Hya) brown dwarf.  As noted
by \citet{hall01}, analogies to the active late-M dwarfs with persistent 
emission (such as LHS 2397a) may also be appropriate.  The correlations
found by \citet{gizis00} would then suggest that  
2MASSI J1315309-264951 may be a relatively
old L dwarf (perhaps even burning hydrogen!).  I note that \citet{magmdwarfs}
have presented nonstandard interior models with strong magnetic fields in 
which magnetically active L dwarfs might actually be stars.   
The possible proper motion
noted by Hall would be inconsistent with TW Hya membership and would suggest
a relatively old kinematic age.  
Further study -- particularly astrometry -- of 
this intriguing L dwarf is needed to understand its evolutionary
status.  If, despite the arguments already given, this L dwarf is 
a TWA member, comparison of 
\citet{dusty00}'s model calculations to its $K_s$ magnitude and
effective temperature would suggest a mass of $\sim 0.013 M_\odot$, near the
deuterium burning limit.  

{\bf Kelu-1}:  The TWA search criteria selected the well-known L dwarf 
Kelu-1 \citep{kelu}, which is known to lie above 
other dwarfs in the HR diagram, but whose 
astrometric distance and motion \citep{dahn} is inconsistent with 
the TW Hya association. 

{\bf 2MASSW J1155395-372735}:  The L2 dwarf, like Kelu-1, shows 
weak H$\alpha$ emission.  The emission was present with similar
strength on two different nights.  It is brighter than Kelu-1
and may serve as a useful L dwarf for further study.  

\subsection{The Mass Function}

There are five known TW Hya members in the range 0.5 to 0.9 $M_\odot$
and 10 in the range 0.2 to 0.5 $M_\odot$, consistent with a 
power-law mass function ($\frac{dN}{dM} \propto M^{-\alpha}$) 
with slope $\alpha=1$.  Extrapolation of this mass function, which is
consistent with field and cluster studies, predicts
$\sim 4$ brown dwarfs in the range 0.015 to 0.025 $M_\odot$.
The observations of two strong candidate isolated brown dwarf
in this mass range 
are consistent with this distribution.  
(Note that one might also count the wide companion TW Hya 5B.)  
In contrast, a mass function
slope of $\alpha =0$ would predict only 0.1 such brown dwarfs. 
Within the very large uncertainties due to the small sample size,
there is no evidence for an IMF in the TW Hya association that
differs from richer star formation regions 
(typically $\alpha \approx 0.5-1.0$).   Studies of the outskirts of the
TWA area and searches of other associations are needed to build up
useful statistics.  

\subsection{Comments on Field Dwarfs}

Most of the field M and L dwarfs observed in this program 
have properties consistent with those observed in previous
2MASS surveys.  The unusual emission of the  L dwarf was discussed
in Section~\ref{twhya_ldwarf}.  Two other systems deserve special
discussion.  

{\bf 2MASSW J1004392-333518}:
This L4 dwarf, discovered in the TWA search, is 7.4 arcsec east and
9.6 arcsec south of the high proper motion star LHS 5166.   
The relationships in \citet{k00} predict $M_J = 12.940$ and
$M_K=11.326$ for an L4 dwarf, both consistent with a distance of 21 pc.
At this distance, the separation of the two components is
250 A.U.   No proper motion is available for the L dwarf
since it is not detected on the sky survey plates, so 
companionship is {\it not} confirmed.
The CTIO 4-meter spectrum reveals that LHS 5166 is
an M4 dwarf with H$\alpha$ emission  of 2.6\AA.  The 2MASS photometry
is $J=9.859$, $H=9.336$, and $K_s = 9.035$.  This spectral type is
at the position of a kink in the main sequence: at LHS 5166's
TiO5 index value of 0.39, the main sequence covers the range 
$6.5 < M_K < 8.5$  \citep{gr96}.  The L dwarf's distance estimate
implies $M_K = 7.43$ (if the components are single), which
is consistent with the large uncertainties for M4 dwarfs.  
At spectral type M4, 31\% of stars have $H\alpha$ emission \citep{pmsu2};
if star formation is constant over 10 Gyr, then the system is 
probably no older than $\sim 3.1$ billion years old.   

{\bf 2MASSW J1510478-281817}:
An M9 dwarf, 2MASSW J151047-281823, is 3.3 arcseconds west and
5.9 arcseconds south of the M8 dwarf 2MASSW J151047-281817.
The proximity suggests that this may be a binary system, 
although the differences in magnitude ($\Delta J = 1.16$,
$\Delta H = 1.22$, $\Delta K_s = 1.07$) are surpisingly
large given the one class difference in spectral type. 
The M8 has $H\alpha$ emission ($EW = 2.5$\AA) while the
M9 does not, but this is consistent with fall-off in activity
with spectral type \citep{gizis00} and does not rule out
binarity.  The M8 has a photometric distance estimate of 
20 parsecs ($M_K = 10.19$, assuming it is on the main sequence) 
while the M9's estimate is 30 parcsecs
($M_K = 10.39$); the corresponding
orbital separation would then be 135 or 200 A.U.  If the
M8 is an equal luminosity double then the discrepency would be explained.  
Approximately 20\% of 2MASS-selected late-M dwarfs
are doubles when imaged by Hubble Space Telescope (Gizis et al., in prep.), so
this scenario is plausible.  
The USNO scans do not have a clear detection of the M9 dwarf,
and therefore do not reveal whether the proper motions are
consistent.  Additional data are required to show that this a 
physically related system rather than a chance alignment.  
If confirmed by common proper motions and radial velocities, this
would be widest known system composed only of M8 or later components.

\section{Summary}

A 2MASS-based search for isolated TWA brown dwarfs has found two
late-M brown dwarfs which are candidate TWA members.  
None of the L dwarfs observed appear to be TWA members.  
These observations are consistent with the substellar mass
function seen in richer star formation regions.  One young M8 dwarf
shows very strong emission features and is an ideal candidate
for more detailed study of this rare phenomenon; a number of other
dwarfs identified in this search are also unusual and deserve additional 
follow-up.  

This pilot study has confirmed that 2MASS is useful for identifying
young brown dwarf members of nearby associations. 
Extension of this study to either higher mass brown dwarfs of 
spectral type M6-M7
and very-low-mass brown dwarfs of spectral type T 
(below the deuterium burning limit) is practical but
requires additional imaging and spectroscopy.  

\acknowledgments

I thank the 2MASS Rare Objects Team, including Neill Reid,
Jim Liebert, Dave Monet, Davy Kirkpatrick, and Patrick Lowrance 
for many useful discussions.    
This publication makes use of data products from 2MASS, 
which is a joint project of the
University of Massachusetts and IPAC/Caltech, funded by NASA and NSF.
J.E.G. was a Guest User, Canadian Astronomy Data Centre, 
which is operated by the Herzberg Institute of Astrophysics, 
National Research Council of Canada. 
This research has made use of the USNOFS Image and Catalogue Archive
operated by the United States Naval Observatory, Flagstaff Station
(http://www.nofs.navy.mil/data/fchpix/).  

\clearpage

\begin{deluxetable}{lrrrlr}
\tablewidth{0pc}
\tablenum{1}
\tablecaption{TWA Targets}
\tablehead{
\colhead{Name} &
\colhead{J} &
\colhead{H} &
\colhead{K$_s$} & 
\colhead{Sp.} &
\colhead{H$\alpha$ EW}
}
\startdata
2MASSW J1004392-333518  & 14.505 & 13.493 & 12.932  & L4 & \nodata \\ 
2MASSW J1012065-304926  & 12.235 & 11.620 & 11.139  & M6 & 5.6 \\ 
2MASSW J1013426-275958  & 12.271 & 11.643 & 11.261  & M5.0 & 6.6 \\ 
2MASSW J1018588-290953  & 14.218 & 13.440 & 12.816  & L1 & \nodata \\ 
2MASSW J1032136-420856  & 12.855 & 12.241 & 11.834  & M7 & 11.8 \\ 
2MASSW J1036530-344138  & 15.635 & 14.462 & 13.806  & L6 & \nodata \\ 
2MASSW J1039183-411033  & 11.962 & 11.337 & 10.941  & M6 & 4.4 \\ 
2MASSW J1045171-260724  & 12.818 & 12.138 & 11.611  & M8 & 6.0 \\ 
2MASSW J1046040-304842  & 12.566 & 11.897 & 11.565  & M5.5 & 10.8 \\ 
2MASSW J1052012-501444  & 13.325 & 12.609 & 12.174  & M8 & 4.7 \\ 
2MASSW J1117540-515302  & 10.665 & 9.766 & 9.279  & gM & \nodata \\ 
2MASSW J1119195-403813  & 10.395 & 9.575 & 9.162  & gM & \nodata \\ 
2MASSW J1122362-391605  & 15.717 & 14.689 & 13.898  & L3 & \nodata \\ 
2MASSW J1132137-264724  & 14.874 & 14.169 & 12.960  & AGN & 13.1 \\ 
2MASSW J1135238-493527  & 15.400 & 14.485 & 13.692  & AGN & \nodata \\ 
2MASSW J1139511-315921  & 12.673 & 11.990 & 11.490  & M8\tablenotemark{a} & 10.0 \\ 
2MASSW J1141423-334133  & 13.854 & 12.778 & 12.177  & C & \nodata \\ 
2MASSW J1144249-430253  & 12.214 & 11.579 & 11.180  & M6 & \nodata \\ 
2MASSW J1148259-380927  & 11.154 & 10.508 & 10.127  & M5.0 & \nodata \\ 
2MASSW J1148542-254440  & 13.408 & 12.679 & 12.172  & M8 & 17.8 \\ 
2MASSW J1153063-250544  & 12.894 & 12.221 & 11.869  & M5.0 & 4.4 \\ 
2MASSW J1155395-372735  & 12.830 & 12.044 & 11.437  & L2 & 2.5 \\ 
2MASSW J1201544-293248  & 12.823 & 12.239 & 11.802  & M6 & 4.9 \\ 
2MASSW J1202410-445538  & 11.964 & 11.015 & 10.708  & gM & \nodata \\ 
2MASSW J1205527-385451  & 12.477 & 11.883 & 11.467  & M5.0 & \nodata \\ 
2MASSW J1206235-401708  & 13.268 & 12.619 & 12.189  & M8 & 1.4 \\ 
2MASSW J1207334-393254  & 12.982 & 12.396 & 11.959  & M8\tablenotemark{a} & 300 \\ 
2MASSW J1208247-420400  & 11.802 & 10.900 & 10.572  & gM & \nodata \\ 
2MASSI J1211024-261409  & 15.267 & 14.349 & 13.163  & QSO & \nodata \\ 
2MASSW J1223135-492544  & 12.324 & 11.667 & 11.317  & M5.0 & \nodata \\ 
2MASSW J1229430-352302  & 11.985 & 11.079 & 10.817  & gM & \nodata \\ 
2MASSW J1238292-405608  & 15.379 & 13.914 & 12.734  & C & \nodata \\ 
2MASSW J1241080-384312  & 11.470 & 10.833 & 10.447  & M5.0 & 10.8 \\ 
2MASSW J1241552-315259  & 12.720 & 12.029 & 11.639  & M6.5 & 7.7 \\ 
Kelu-1  & 13.417 & 12.387 & 11.726  & L2 & 1 \\ 
2MASSI J1315309-264951  & 15.184 & 14.055 & 13.457  & L5 & 97 \\ 
2MASSW J1324055-350806  & 13.384 & 12.669 & 12.270  & M6 & \nodata \\ 
2MASSW J1325580-362035  & 12.363 & 11.734 & 11.321  & M5.0 & 8.3 \\ 
2MASSW J1326201-272937  & 15.835 & 14.717 & 13.841  & L5 & \nodata \\ 
2MASSI J1326401-421912  & 11.697 & 10.767 & 10.484  & gM & \nodata \\ 
2MASSW J1329019-414713  & 13.657 & 12.817 & 12.339  & M9 & \nodata \\ 
\enddata
\tablenotetext{a}{Low surface gravity (see Section 3.1)}
\end{deluxetable}

\begin{table}
\dummytable\label{table-twhya}
\end{table}

\begin{deluxetable}{lrrrrrlrc}
\tablewidth{0pc}
\tablenum{2}
\tablecaption{Field Targets}
\tablehead{
\colhead{Name} &
\colhead{J} &
\colhead{H} &
\colhead{K$_s$} & 
\colhead{Sp.} &
\colhead{H$\alpha$ EW}
}
\startdata
2MASSI J1003191-010507  & 12.352 & 11.685 & 11.267  & M7 & 16.4 \\ 
2MASSW J1045171-260724  & 12.818 & 12.138 & 11.611  & M8 & 6.0 \\ 
2MASSI J1045240-014957  & 13.129 & 12.370 & 11.810  & L1 & \nodata \\
DENIS-P J104814.7-395606.1 & 9.551 & 8.915 & 8.452 & M8 & 2.0 \\ 
2MASSW J1054119-850502  & 12.716 & 12.065 & 11.657  & M8 & \nodata \\ 
2MASSI J1224522-123835  & 12.564 & 11.831 & 11.371  & M8 & 11.3 \\ 
2MASSI J1309218-233035  & 11.769 & 11.087 & 10.666  & M7 & 6.7 \\ 
2MASSI J1359206-302339  & 14.577 & 13.072 & 11.798  & C & \nodata \\ 
2MASSW J1420544-361322  & 11.484 & 10.853 & 10.406  & M7 & 7.0 \\ 
2MASSI J1421187-161820  & 12.769 & 12.109 & 11.675  & M8 & \nodata \\ 
2MASSI J1504162-235556  & 12.025 & 11.389 & 11.031  & M7 & 5.5 \\ 
2MASSW J1510168-024107  & 12.625 & 11.838 & 11.371  & M8 & 4.2 \\ 
2MASSW J1510476-281823  & 14.011 & 13.332 & 12.767  & M9 & \nodata \\ 
2MASSW J1510478-281817  & 12.848 & 12.114 & 11.693  & M8 & 2 \\ 
2MASSW J1515110-133227  & 12.594 & 11.516 & 10.785  & C & \nodata \\ 
2MASSI J1534570-141848  & 11.390 & 10.731 & 10.311  & M8 & 2.6 \\ 
2MASSW J1555157-095605  & 12.531 & 11.993 & 11.451  & L1 & 3.1 \\ 
2MASSW J1558187-205419  & 12.037 & 11.406 & 10.967  & M5.5 & 19.8 \\ 
2MASSW J1607312-044209  & 11.882 & 11.176 & 10.717  & M8 & 7.9 \\ 
2MASSW J1611453-192813  & 12.871 & 12.187 & 11.814  & M5.0 & 10.5 \\ 
2MASSW J1614232-125046  & 13.034 & 12.435 & 11.983  & M7 & 5.0 \\ 
2MASSW J1632588-063148  & 12.757 & 12.044 & 11.628  & M7 & 6.6 \\ 
2MASSW J1635369-114710  & 12.913 & 12.229 & 11.817  & M3 & \nodata \\ 
2MASSW J1645221-131951  & 12.467 & 11.723 & 11.195  & L1.5 & \nodata \\ 
2MASSI J1652277-160738  & 13.130 & 12.130 & 11.525  & C & 1.1 \\ 
2MASSW J1655379+045430  & 12.364 & 11.759 & 11.282  & M8 & 2.8 \\ 
2MASSW J1707234-055824  & 12.062 & 11.274 & 10.710  & M9 & 0.4 \\ 
2MASSW J1832238-443508  & 12.978 & 12.370 & 11.918  & M6 & 7.1 \\ 
2MASSW J1930274-194349  & 12.390 & 11.731 & 11.283  & M6.5 & 6.0 \\ 
\enddata
\end{deluxetable}

\begin{table}
\dummytable\label{table-field}
\end{table}

\begin{deluxetable}{lrrrrrr}
\tablewidth{0pc}
\tablenum{3}
\tablecaption{Spectral Indices}
\tablehead{
\colhead{Name} &
\colhead{TiO5} &
\colhead{PC3} &
\colhead{CrH} & 
\colhead{TiO-b} &
\colhead{VO-a} &
\colhead{VO-b} 
}
\startdata
Gl 300  &  0.40 &  1.13 &  0.96 &  1.23 &  0.99 &  1.03 \\ 
Gl 369  &  0.74 &  1.01 &  1.00 &  1.04 &  0.99 &  0.99 \\
Gl 831A  &  0.37 &  1.16 &  0.97 &  1.28 &  1.00 &  1.04 \\ 
Kelu-1  &  1.05 &  2.72 &  1.55 &  1.24 &  1.03 &  1.15 \\ 
LHS 2065  &  0.51 &  2.23 &  1.07 &  2.02 &  1.20 &  1.51 \\ 
LHS 2397a  &  0.34 &  2.10 &  1.10 &  2.11 &  1.15 &  1.48 \\ 
LHS 5166  &  0.39 &  1.16 &  0.99 &  1.21 &  1.00 &  1.01 \\ 
TVLM 868-53850  &  0.27 &  1.33 &  0.99 &  1.50 &  1.03 &  1.13 \\ 
VB 8  &  0.17 &  1.67 &  1.04 &  1.88 &  1.08 &  1.27 \\ 
2MASSW J1003191-010507  &  0.24 &  1.81 &  1.10 &  1.99 &  1.09 &  1.34 \\ 
2MASSW J1004392-333518    &  1.11 &  4.38 &  1.91 &  1.27 &  0.82 &  1.04 \\
2MASSW J1012065-304926  &  0.23 &  1.57 &  0.99 &  1.84 &  1.08 &  1.29 \\ 
2MASSW J1013426-275958  &  0.30 &  1.33 &  0.98 &  1.47 &  1.01 &  1.11 \\ 
2MASSW J1018588-290953    &  0.86 &  2.65 &  1.35 &  1.49 &  1.08 &  1.25 \\
2MASSW J1032136-420856  &  0.22 &  1.88 &  1.12 &  2.01 &  1.09 &  1.34 \\ 
2MASSW J1036530-344138  &  0.84 & 65.57 &  1.83 &  0.91 &  0.33 &  0.98 \\ 
2MASSW J1039183-411033  &  0.26 &  1.47 &  1.02 &  1.59 &  1.04 &  1.15 \\ 
2MASSW J1045171-260724  &  0.26 &  1.96 &  1.06 &  2.22 &  1.19 &  1.51 \\ 
2MASSW J1045240-014957    &  0.99 &  2.71 &  1.35 &  1.52 &  1.10 &  1.29 \\
2MASSW J1046040-304842  &  0.24 &  1.39 &  1.00 &  1.54 &  1.04 &  1.13 \\ 
2MASSW J1052012-501444  &  0.23 &  2.06 &  1.11 &  2.25 &  1.17 &  1.51 \\ 
2MASSW J1054119-850502  &  0.29 &  2.16 &  1.08 &  2.27 &  1.15 &  1.51 \\ 
2MASSW J1117540-515302  &  0.30 &  1.15 &  0.90 &  1.49 &  0.97 &  1.05 \\ 
2MASSW J1119195-403813  &  0.32 &  1.03 &  0.90 &  1.43 &  0.95 &  1.03 \\ 
2MASSW J1122362-391605  &  0.81 &  4.39 &  1.65 &  1.21 &  1.14 &  1.09 \\ 
2MASSW J1132137-264724  &  0.99 &  0.87 &  0.99 &  1.02 &  0.98 &  1.00 \\ 
2MASSW J1135238-493527  &  1.33 &  1.03 &  1.00 &  1.01 &  0.98 &  0.97 \\ 
2MASSW J1139511-315921  &  0.26 &  2.02 &  0.91 &  2.56 &  1.33 &  1.71 \\ 
2MASSW J1141423-334133  &  0.60 &  0.79 &  0.90 &  0.88 &  1.11 &  1.10 \\ 
2MASSW J1144249-430253  &  0.22 &  1.46 &  0.96 &  1.65 &  1.05 &  1.19 \\ 
2MASSW J1148259-380927  &  0.31 &  1.29 &  0.94 &  1.46 &  1.02 &  1.12 \\ 
2MASSW J1148542-254440  &  0.43 &  2.14 &  1.04 &  2.25 &  1.16 &  1.52 \\ 
2MASSW J1153063-250544  &  0.26 &  1.35 &  1.01 &  1.42 &  1.02 &  1.08 \\ 
2MASSW J1155395-372735    &  0.98 &  3.00 &  1.62 &  1.35 &  1.03 &  1.22 \\
2MASSW J1201544-293248  &  0.23 &  1.51 &  0.97 &  1.82 &  1.07 &  1.26 \\ 
2MASSW J1202410-445538  &  0.71 &  1.03 &  0.99 &  1.04 &  0.98 &  1.00 \\ 
2MASSW J1205527-385451  &  0.25 &  1.34 &  0.96 &  1.54 &  1.04 &  1.14 \\ 
2MASSW J1206235-401708    &  0.32 &  1.90 &  1.00 &  2.23 &  1.17 &  1.54 \\
2MASSW J1207334-393254    &  0.29 &  1.80 &  0.92 &  2.38 &  1.25 &  1.60 \\
2MASSW J1208247-420400  &  0.52 &  1.03 &  0.97 &  1.12 &  0.97 &  0.99 \\ 
2MASSW J1211024-261409  &  1.27 &  0.84 &  0.99 &  0.97 &  1.03 &  1.06 \\ 
2MASSW J1223135-492544  &  0.26 &  1.33 &  0.93 &  1.57 &  1.05 &  1.17 \\ 
2MASSW J1224522-123835  &  0.40 &  1.96 &  1.02 &  2.19 &  1.20 &  1.55 \\ 
2MASSW J1229430-352302  &  0.26 &  1.06 &  0.85 &  1.73 &  0.95 &  1.09 \\ 
2MASSW J1235366-411203  &  1.07 &  1.01 &  1.02 &  1.00 &  1.01 &  0.99 \\ 
2MASSW J1238292-405608  &  0.93 &  1.08 &  1.04 &  0.98 &  1.02 &  1.04 \\ 
2MASSW J1241080-384312  &  0.29 &  1.41 &  0.99 &  1.62 &  1.05 &  1.19 \\ 
2MASSW J1241552-315259    &  0.23 &  1.58 &  0.98 &  1.92 &  1.11 &  1.34 \\
2MASSW J1309218-233035  &  0.24 &  1.89 &  1.02 &  2.11 &  1.14 &  1.45 \\ 
2MASSW J1315309-264951    &  1.33 &  9.12 &  2.04 &  1.06 &  0.79 &  1.07 \\
2MASSW J1324055-350806  &  0.23 &  1.49 &  0.96 &  1.84 &  1.12 &  1.33 \\ 
2MASSW J1325580-362035  &  0.29 &  1.26 &  0.95 &  1.51 &  1.02 &  1.13 \\ 
2MASSW J1326201-272937  &  0.86 & 17.23 &  1.89 &  1.06 &  0.35 &  1.01 \\ 
2MASSW J1326401-421912  &  0.35 &  1.07 &  0.92 &  1.36 &  0.97 &  1.01 \\ 
2MASSW J1329019-414713    &  0.69 &  2.12 &  1.05 &  1.69 &  1.20 &  1.40 \\
2MASSW J1359206-302339  &  0.77 &  0.97 &  1.01 &  1.01 &  1.10 &  0.95 \\ 
2MASSW J1420544-361322  &  0.20 &  1.75 &  1.08 &  2.00 &  1.10 &  1.33 \\ 
2MASSW J1421187-161820  &  0.26 &  1.94 &  1.16 &  1.95 &  1.07 &  1.32 \\ 
2MASSW J1504162-235556  &  0.21 &  1.69 &  1.04 &  1.95 &  1.11 &  1.33 \\ 
2MASSW J1507477-162738    &  1.12 &  7.50 &  2.14 &  1.11 &  0.89 &  1.01 \\
2MASSW J1510168-024107  &  0.41 &  2.11 &  1.15 &  2.12 &  1.18 &  1.51 \\ 
2MASSW J1510478-281817    &  0.30 &  2.01 &  1.00 &  2.46 &  1.30 &  1.71 \\
2MASSW J1515110-133227  &  0.52 &  0.83 &  0.92 &  0.91 &  1.11 &  1.03 \\ 
2MASSW J1534570-141848  &  0.19 &  2.05 &  1.12 &  2.20 &  1.10 &  1.38 \\ 
2MASSW J1555157-095605  &  0.74 &  3.31 &  1.44 &  1.68 &  1.10 &  1.42 \\ 
2MASSW J1558187-205419  &  0.26 &  1.39 &  0.90 &  1.77 &  1.05 &  1.23 \\ 
2MASSW J1607312-044209  &  0.27 &  1.94 &  1.04 &  2.13 &  1.18 &  1.48 \\ 
2MASSW J1611453-192813  &  0.28 &  1.33 &  0.92 &  1.56 &  1.01 &  1.14 \\ 
2MASSW J1614232-125046  &  0.17 &  1.67 &  1.04 &  1.91 &  1.10 &  1.30 \\ 
2MASSW J1632588-063148  &  0.25 &  1.99 &  1.11 &  2.17 &  1.14 &  1.49 \\ 
2MASSW J1635369-114710  &  0.47 &  1.17 &  0.98 &  1.14 &  0.99 &  1.00 \\ 
2MASSW J1645221-131951    &  0.87 &  2.89 &  1.45 &  1.46 &  1.07 &  1.27 \\
2MASSW J1652277-160738  &  0.53 &  0.92 &  0.92 &  0.90 &  1.14 &  0.95 \\ 
2MASSW J1655379+045430    &  0.23 &  2.19 &  1.12 &  2.26 &  1.12 &  1.44 \\
2MASSW J1707234-055824    &  0.60 &  2.43 &  1.19 &  1.91 &  1.18 &  1.48 \\
2MASSW J1832238-443508  &  0.25 &  1.47 &  1.00 &  1.71 &  1.06 &  1.22 \\ 
2MASSW J1930274-194349    &  0.23 &  1.65 &  1.04 &  1.92 &  1.10 &  1.31 \\
\enddata
\end{deluxetable}

\begin{table}
\dummytable\label{table-indices}
\end{table}


\end{document}